\documentclass[11pt]{article}

\usepackage[T1]{fontenc}
\usepackage[utf8]{inputenc}
\usepackage{lmodern}
\usepackage[margin=0.8in]{geometry}
\usepackage{microtype}
\usepackage{setspace}
\usepackage{parskip}
\usepackage{enumitem}
\usepackage{booktabs}
\usepackage{threeparttable}
\usepackage{array}
\usepackage{tabularx}
\usepackage{longtable}
\usepackage{graphicx}
\usepackage{float}
\usepackage{caption}
\usepackage{subcaption}
\usepackage{xcolor}
\usepackage{tikz}
\usetikzlibrary{arrows.meta,positioning,shapes.geometric,calc}
\usepackage{hyperref}
\usepackage{placeins}
\usepackage{cleveref}
\usepackage[round,authoryear]{natbib}
\usepackage{authblk}

\usepackage{etoolbox}
\usepackage{csvsimple} 
\usepackage{siunitx} 
\hypersetup{
  colorlinks=true,
  linkcolor=blue!50!black,
  citecolor=blue!50!black,
  urlcolor=blue!50!black,
  pdftitle={Online and Survey Representations of Well-Being: Coherence and Cross-Source Correspondence}
}

\newcommand{\maybegraphic}[3][0.95\textwidth]{%
  \IfFileExists{#2}{\includegraphics[width=#1]{#2}}{%
    \fbox{\parbox[c][2.3in][c]{0.9\textwidth}{\centering
    Missing figure: \texttt{#2}\\[0.5em]#3}}}}

\title{Online and Survey Representations of Well-Being: Coherence and Cross-Source Correspondence}

\author[1]{Stefano M.\ Iacus}
\author[2]{Giuseppe Porro}
\author[3,4]{Haodong Qi\thanks{Corresponding author: \href{mailto:haodong.qi@mau.se}{haodong.qi@mau.se}}}

\affil[1]{Institute for Quantitative Social Science, Harvard University, Cambridge, MA 02138, USA}
\affil[2]{Department of Law, Economics and Culture, University of Insubria, Como, Italy}
\affil[3]{Stockholm University Demography Unit, Stockholm, Sweden}
\affil[4]{Department of Global Political Studies, Malm\"o University,  Malm\"o, Sweden}

\begin{document}
\maketitle

\begin{abstract}
Digital traces may complement population surveys in the measurement of
well-being, but the two sources do not necessarily capture identical constructs. We
compare the Human Flourishing Geographic Index (HFGI) with World Values Survey
and European Social Survey measures across three domains using structural
equation models. The results show strong within-source coherence and
construct-specific cross-source correspondence. Cross-source
convergence varies across domains; the observed pattern is consistent with
stronger correspondence where the underlying indicators are more closely
aligned conceptually. Near-perfect
agreement would be unexpected, since surveys elicit standardized self-reports,
whereas HFGI captures 
naturally occurring public expressions. HFGI should
therefore be viewed as a distinct but empirically connected measurement system
that complements survey-based evidence on well-being.
\end{abstract}


\section*{Acknowledgments}
Computational resources were provided by the National Academic Infrastructure for Supercomputing in Sweden (NAISS), funded by the Swedish Research Council. Funding support from the Belmont Forum, the Swedish Research Council Vetenskapsrådet (grant agreement 2022–06012-3) and U.S. National Science Foundation (Award Number: 2310908) is gratefully acknowledged.

\newpage

\section{Introduction}

National well-being has become a central component of the assessment of social
progress and an increasingly important complement to conventional economic
indicators. GDP and related measures remain indispensable for describing
production and material resources, but they do not directly capture how people
evaluate their lives, experience positive or negative emotions, or perceive
their capacity to act and influence their circumstances
\citep{diener2009well-being}. This has motivated the broader ``Beyond GDP''
agenda and the development of national statistical frameworks in which
subjective well-being is treated as a relevant dimension of economic and
social performance
\citep{stiglitz2009report,oecd2013guidelines,oecd2025guidelines,
helliwell2025world}. Well-being research has consequently relied heavily on
population surveys measuring constructs such as happiness, life satisfaction,
and perceived control. These measures remain essential for comparing how
people assess their lives across countries and over time, although the
distinction between evaluative and affective well-being must be preserved
\citep{diener1984subjective,deaton2013happiness}.

Survey-based measurement nevertheless has important practical limitations.
Cross-national surveys are costly and are generally conducted in periodic
waves, while national estimates may refer to different years and require
complex harmonisation procedures. They therefore offer relatively sparse
observations of a phenomenon that may change in response to economic,
political, and social shocks. The increasing availability of digital traces
offers a potentially valuable complement. Social-media data contain large
volumes of geographically referenced language produced continuously and at
relatively low marginal cost, making it possible to study social processes at
a temporal and spatial resolution that traditional surveys rarely provide
\citep{salganik2017bit,iacusporro2021subjective}. At the same time, the
ready-made character of digital trace data does not remove the need for
sampling, validation, and measurement design: the advantages of scale and
timeliness coexist with problems of selection, platform dependence, changing
data-generating processes, and the risk that large datasets create a
misleading appearance of representativeness
\citep{lazer2014parable,olteanu2019social,hargittai2020potential}.

The Human Flourishing Geographic Index (HFGI) was developed in this context.
Its conceptual basis is multidimensional flourishing, rather than a single
affective measure of happiness, and is connected to the flourishing framework
developed by the Harvard Human Flourishing Program
\citep{vanderweele2017promotion,rule2024measurement}. The HFGI is constructed from geolocated social-media posts contained in the Harvard CGA Geotweet Archive \citep{lewis16}. The
first systematic HFGI dataset and applications, however, concentrated on the
approximately 2.6 billion geolocated U.S. posts available in the repository, using fine-tuned language models to construct multidimensional
indicators of human flourishing \citep{iacus2026hfgi}. These initial studies
showed, in brief, that HFGI can document county-level differences between
evaluative and affective well-being, identify rural--urban variation in
job-satisfaction expressions, and provide a fine-grained geographic and
temporal resource for well-being research
\citep{iacus2026twoamericas,iacus2026jobsatisfaction, IacusTransGis2026}. Related regional work
further suggests that the association between territorial disadvantage and
subjective well-being changes with geographical scale and temporal dynamics
\citep{iacus2026leftbehindness}.

These findings suggest that online indicators are substantively informative,
but they do not establish that they measure national well-being in the same way
as surveys. The fact that an online indicator displays theoretically plausible
associations, or that several HFGI dimensions are internally related, does not
demonstrate that its cross-national variation corresponds to variation in
survey-reported well-being.

Digital traces may capture forms of well-being expression that are related to,
but not identical with, the evaluations elicited by conventional surveys.
Establishing this relationship is therefore an empirical question rather than
an assumption. The present study extends the empirical scope of
HFGI beyond its initial U.S.-focused applications by examining its
cross-national correspondence with two external survey sources, the World
Values Survey (WVS) and the European Social Survey (ESS).

The issue matters for both the interpretation and the practical
use of digital well-being measures. If online and survey indicators recover
similar cross-national structure, digital traces may provide complementary
evidence alongside conventional surveys. Where correspondence is weaker, the
two sources may instead capture different dimensions of a broader construct.
We therefore examine three domains for which meaningful cross-source
comparisons can be constructed: mental and subjective well-being, spiritual
and religious expression, and physical well-being. Rather than assuming that
similarly labelled indicators are equivalent, we ask whether each source forms
a coherent measurement system and the extent to which the resulting latent
constructs correspond across countries.

\subsection{Measurement problem}

Comparing online and survey-based well-being measures raises a measurement
problem that cannot be resolved simply by correlating indicators with similar
labels. The same construct may be observed through different measurement
systems, while the same measurement system may capture several related but
distinct aspects of well-being. The relevant question is therefore not whether
HFGI and survey indicators use comparable terminology, but whether they
represent the same cross-national variation in an underlying well-being
construct.

The two sources differ first in their effective populations. Survey indicators
are derived from respondents selected through a sampling design and are
commonly weighted to approximate the target population. HFGI indicators, by
contrast, are based on geolocated social-media posts produced by a self-selected
population of users. Geotagged users constitute a small and socially
distinctive subset of the broader population, and their composition may vary
across countries according to age, education, income, digital access, platform
use, language, and cultural norms
\citep{mislove2011,malik2015,sloan2015geotagging,mellon2017twitter}. HFGI
therefore measures the geographically aggregated propensity to express
well-being-related content online, rather than the prevalence of a given
well-being state in the population as a whole \citep{iacus2026hfgi}. Large
numbers of posts may reduce sampling variability within the observed digital
population, but they cannot by themselves eliminate selection into that
population.

The two sources also differ in how information is elicited. Survey respondents
are explicitly asked to evaluate their lives or to report their happiness and
perceived control in response to standardized questions. Social-media users are
not generally answering a well-being questionnaire. They communicate in a
public or semi-public environment for purposes that may include social
interaction, self-presentation, information sharing, entertainment, or
emotional expression. The language classified by HFGI is consequently shaped
not only by users' well-being but also by the communicative situation, the
intended audience, platform conventions, and the likelihood that a particular
experience will be expressed publicly. This distinction is central to the
interpretation of social-media well-being indicators: they measure expressed
discourse, not necessarily the latent prevalence of the corresponding
experience \citep{iacusporro2022listen,pavalanathan2015confounds}. Recent work
using a multi-trait, multi-method framework confirms that digital-trace and
survey measures can contain distinct sources of measurement error even when
they are intended to measure similar traits \citep{cernat2025measurement}.

A further difficulty concerns the conceptual correspondence
between the indicators. Happiness, life satisfaction, inner peace, depression,
anxiety, and perceived choice or control capture related but non-identical
aspects of well-being. Life satisfaction is generally interpreted as a
reflective evaluation of one's life as a whole, whereas happiness is more
closely associated with affective experience. Inner peace, depression, and
anxiety capture experiential and psychological states, while choice and
control concern perceived agency. The HFGI mental factor therefore combines
positive evaluative and experiential indicators with adverse psychological
indicators, whereas the WVS factor combines happiness, life satisfaction, and
perceived agency. Conceptual proximity is thus compatible with differences in
the breadth and empirical meaning of the two source-specific constructs.

The two sources also differ in their temporal coverage. Although the present
analysis operates at the country-year level, HFGI is constructed from
expressions observed continuously throughout the year and subsequently
aggregated. Survey measures, by contrast, are typically collected within a
fieldwork window of one or two months. Their estimates may therefore be
influenced by the season of data collection or by contemporaneous events, such
as economic shocks, geopolitical crises, or public-health emergencies. HFGI
and survey indicators may consequently refer to the same calendar year while
capturing different temporal mixtures of experiences. Continuous digital
observation offers a potentially richer representation of annual well-being,
but it also preserves short-term fluctuations that must be distinguished from
the underlying annual level.

Aggregation introduces another layer of non-equivalence. HFGI country-year
indicators are constructed by aggregating classified posts, with the
contribution of each dimension depending on the volume and salience of
relevant expressions. Survey country-level measures are typically averages or
weighted distributions of individual responses. These operations do not
produce equivalent summaries: the former describe patterns of expressed
language among observed users, while the latter describe reported attitudes
among sampled respondents. Approximate temporal matching may further increase
the discrepancy. Moreover, expressed well-being may respond more rapidly than
reflective evaluations to economic, political, or public-health shocks.

These differences do not imply that either source is inherently invalid.
Survey measures are themselves affected by response styles, question wording,
translation, non-response, and cross-national differences in the
interpretation of response scales. Establishing comparability therefore
requires attention to measurement invariance and to the possibility that
apparently common scales function differently across cultural contexts
\citep{jaidka2020estimating,oecd2013guidelines,krumpal2013determinants}.
Digital measures add further complexities, including classifier error, language
coverage, platform-specific conventions, changes in user composition, and the
distinction between the salience of an expression and the prevalence of an
experience. These are not merely technical inconveniences: social-data
research has documented systematic biases, omitted voices, and ethical
boundaries that can affect substantive conclusions
\citep{olteanu2019social,hargittai2020potential}. Fine-tuned open-source models
improve scalability, transparency, and reproducibility, but automated
classification remains a measurement stage whose errors may vary across
languages, topics, and countries \citep{carammia2026rethinking}.

For these reasons, a simple correlation between individual HFGI and survey
indicators would be difficult to interpret. A high correlation could reflect
shared country-level structure without demonstrating that the two sources
measure the same latent construct. Conversely, a modest correlation could
arise because the sources capture complementary dimensions of well-being rather
than because one of them is invalid. The relevant distinction is therefore
between \emph{within-source coherence} and \emph{between-source
correspondence}. Indicators may form coherent constructs within both the
online and survey systems while the two constructs remain only partially
aligned.

The present study addresses this problem by estimating separate latent factors
for online-expressed and survey-reported well-being. This design allows each
source to contain meaningful common information while leaving their
cross-source association to be estimated rather than assumed. The objective is
not to determine which source provides the ``true'' measure of well-being, but
to establish whether the two systems can be treated as interchangeable for
cross-national comparison or whether they should instead be interpreted as
complementary representations of national well-being.

\subsection{Contribution and research questions}

In summary, this study contributes to research on national well-being
measurement by distinguishing internal measurement coherence from correspondence
across measurement systems. Using latent-variable models and two independent
survey sources, we examine whether HFGI and survey indicators form coherent
source-specific constructs and how cross-source correspondence varies across
three substantive domains. The analysis therefore treats external validation
as construct-specific rather than as a binary test of agreement between
digital traces and surveys.

Consequently, we address the following research questions:

\begin{enumerate}[label=\arabic*.]
  \item Do HFGI and survey indicators form coherent
  source-specific latent constructs?

  \item To what extent do the latent online-expression and
  survey-report factors correspond across countries?

  \item Does the degree of cross-source correspondence vary
  across substantive domains, and is the observed pattern consistent with
  differences in the conceptual alignment of the underlying indicators?
\end{enumerate}

\section{Conceptual framework}\label{sec:framework}

\subsection{Coherence versus correspondence}
Within-source coherence concerns whether indicators from the same measurement system reflect a common latent construct. Between-source correspondence concerns whether the latent constructs rank countries similarly. Strong coherence within each source need not imply strong correspondence across sources.

\subsection{Conceptual measurement model}
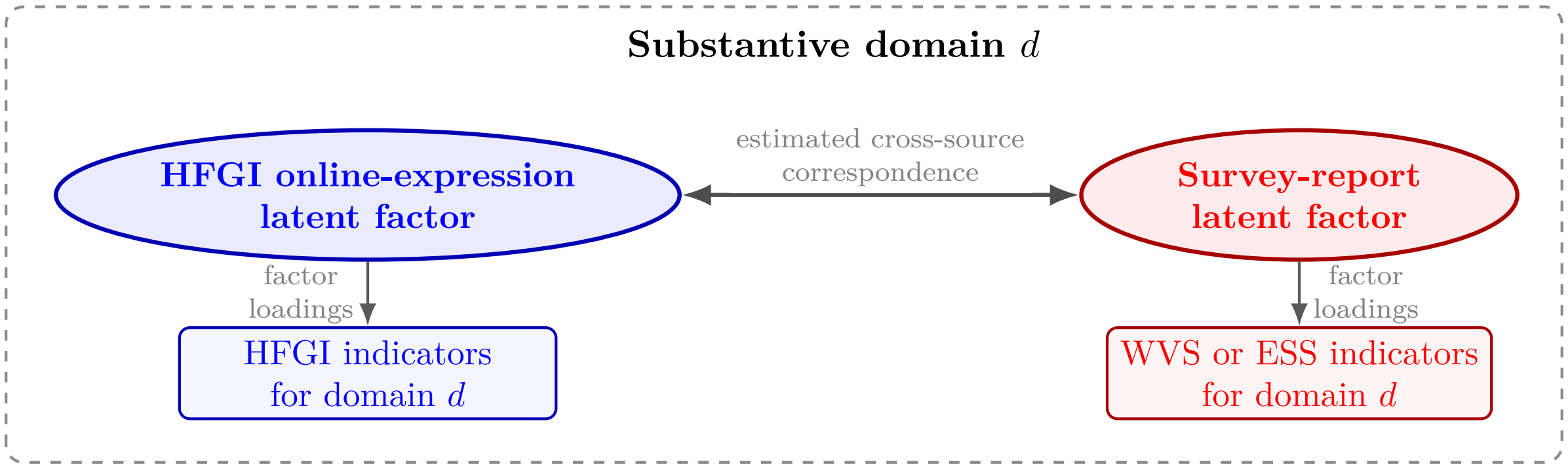
\begin{figure}[H]
\centering

\begin{tikzpicture}[
  latent/.style={
    ellipse,
    very thick,
    minimum width=4.4cm,
    minimum height=1.25cm,
    align=center,
    font=\small\bfseries
  },
  indicators/.style={
    rectangle,
    rounded corners=3pt,
    thick,
    minimum width=3.8cm,
    minimum height=0.9cm,
    align=center,
    font=\small
  },
  measurementpath/.style={
    -{Latex[length=2.3mm]},
    thick,
    draw=black!65
  },
  correspondence/.style={
    <->,
    >=Latex,
    very thick,
    draw=black!70
  }
]

\draw[dashed, rounded corners=5pt, thick, draw=black!45]
  (-8.35,-2.15) rectangle (7.35,2.45);
\node[font=\normalsize\bfseries] at (0,2.08)
  {\textcolor{black}{Substantive domain $d$}};

\node[latent, draw=blue!70!black, fill=blue!8]
  (hfgiFactor) at (-4.7,0.55)
  {\textcolor{blue}{HFGI online-expression}\\\textcolor{blue}{latent factor}};
\node[latent, draw=red!65!black, fill=red!8]
  (surveyFactor) at (4.7,0.55)
  {\textcolor{red}{Survey-report}\\\textcolor{red}{latent factor}};

\node[indicators, draw=blue!70!black, fill=blue!4]
  (hfgiIndicators) at (-4.7,-1.25)
  {\textcolor{blue}{HFGI indicators}\\\textcolor{blue}{for domain $d$}};
\node[indicators, draw=red!65!black, fill=red!4]
  (surveyIndicators) at (4.7,-1.25)
  {\textcolor{red}{WVS or ESS indicators}\\\textcolor{red}{for domain $d$}};

\draw[measurementpath] (hfgiFactor) --
  node[left, font=\scriptsize, align=center]
  {\textcolor{gray}{factor}\\\textcolor{gray}{loadings}}
  (hfgiIndicators);
\draw[measurementpath] (surveyFactor) --
  node[right, font=\scriptsize, align=center]
  {\textcolor{gray}{factor}\\\textcolor{gray}{loadings}}
  (surveyIndicators);

\draw[correspondence] (hfgiFactor) --
  node[above, fill=white, inner sep=2pt, font=\scriptsize, align=center]
  {\textcolor{gray}{estimated cross-source}\\\textcolor{gray}{correspondence}}
  (surveyFactor);

\end{tikzpicture}
\caption{\textcolor{black}{Source-specific latent measurement model applied separately to each substantive
domain: HFGI indicators measure an online-expression latent factor, while WVS
or ESS indicators measure a separate survey-report latent factor. The
double-headed arrow represents the estimated latent correlation between the two
source-specific factors. No common latent factor or cross-source measurement
equivalence is assumed.}}
\label{fig:conceptual-framework-compact}
\end{figure}

For each substantive domain, we use structural equation modeling to estimate two separate latent factors: one underlying the HFGI online-expression indicators and one underlying the survey-report indicators (Figure~\ref{fig:conceptual-framework-compact}). A latent factor represents the common, unobserved construct inferred from the variation shared by its observed indicators. Each factor loading measures the strength and direction of the association between an indicator and its corresponding latent factor. The pattern of loadings therefore describes the degree of \emph{within-source coherence}. We then estimate the correlation between the two source-specific factors to assess \emph{cross-source correspondence}.

\section{Data and methods}\label{sec:data}

{Our objective is to compare HFGI indicators with survey-reported
measures wherever the available survey data permit a substantively meaningful
comparison. No single cross-national survey provides adequate coverage across
all HFGI domains. We therefore combine two complementary survey sources. The
World Values Survey (WVS) provides broad international coverage and a rich set
of indicators of subjective, mental, social, and attitudinal dimensions of
well-being. For physical well-being, however, it contains only a single directly
relevant item, self-rated health. The European Social Survey (ESS), by contrast,
provides multiple physical-health indicators together with sufficiently broad
country coverage and a sufficiently long time span to construct a
country--year panel. We therefore use the WVS for the main cross-domain
analysis and the ESS as a complementary source for testing the correspondence
between HFGI and survey measures in the physical well-being domain.}

\subsection{Human Flourishing Geographic Index}

The social-media indicators used in this study are drawn from the
\textit{Human Flourishing Geographic Index} (HFGI)
\citep{iacus2026hfgi}. The HFGI is conceptually inspired by the framework
developed by Harvard's Human Flourishing Program
\citep{vanderweele2017promotion}, which defines flourishing as a multidimensional
construct encompassing happiness and life satisfaction, mental and physical
health, meaning and purpose, character and virtue, close social relationships,
and material and financial stability. The HFGI extends this framework to
online expression by using fine-tuned large language models
\citep{carammia2026rethinking} to classify approximately 10 billion
georeferenced tweets from the Harvard Center for Geographic Analysis Geotweet
Archive \citep{lewis16}. Full details on indicator construction, model
fine-tuning, accuracy and validation of the classification are provided by \citet{iacus2026hfgi}.

The international HFGI archive available for this study contains 32 indicators
for 61 countries, with annual observations spanning 2011--2023. Coverage is
not completely balanced across countries. Each country--year indicator is
recorded on a dimension-specific scale bounded between $-1$ and $+1$. The data
also report the total number of geolocated tweets in each country--year and the
number classified as valid for each HFGI dimension.
The three domain-specific structural equation models (SEMs) use 10 of the 32 HFGI
indicators: five in the mental and subjective well-being model, two in the
physical-adversity model, and three in the spiritual and religious expression
model. The indicators used in each model are reported in
Table~\ref{tab:hfgi-survey-crosswalk}.

Tweet counts measure the volume of social-media evidence underlying the HFGI
indicators. They do not constitute population sampling weights and should not
be interpreted as reflecting the demographic composition of a country's
population. Potential sources of selection and measurement bias are discussed
in Section~\ref{sec:issues}.

\subsection{World Values Survey Data}

The WVS variables entering the analysis are country--survey-year aggregates
derived from Waves 6 and 7 of the World Values Survey
\citep{inglehart2014wvs6,haerpfer2024wvs7}. Wave 6 officially covers the
2010--2014 fieldwork period, whereas Wave 7 covers 2017--2022. The files used
in this study nevertheless contain national surveys conducted between 2010
and 2016 for Wave 6 and between 2017 and 2023 for Wave 7. Countries are
observed only in the years in which a national WVS survey was conducted;
consequently, the WVS data do not constitute a complete annual panel.

Valid responses were recoded to the interval $[0,1]$ before aggregation.
Binary indicators are represented by country-level proportions, while ordinal
indicators are represented by country-level means of linearly rescaled
response scores. Higher values indicate a greater prevalence or intensity of
the construct named by the indicator.
The two WVS-based domain SEMs use six indicators: three in
the mental and subjective well-being model and three in the spiritual and
religious expression model. The indicators used in each model are reported in
Table~\ref{tab:hfgi-survey-crosswalk}.

The SEM analysis uses only precomputed
country--survey-year aggregates. During construction of these aggregates,
valid recoded responses were weighted using the respondent weights supplied by
the WVS for both  Wave 6 and Wave 7.
Binary indicators therefore represent weighted country-level proportions,
whereas ordinal indicators represent weighted means of the rescaled response
scores. These weights enter only in the construction of the aggregates; the
SEM itself uses neither individual-level observations nor additional survey
weights.

\subsection{European Social Survey Data}
\label{sec:ess-data}

{The ESS complements the WVS specifically for the physical
well-being domain. Whereas the WVS provides only one directly relevant
physical-health item, the ESS provides multiple indicators that can support a
survey-side latent physical construct.} We use country-level aggregates from
ESS Rounds 5--11 \citep{jowell2007}. Two core items define the survey-side
physical-adversity construct: general self-rated health (\texttt{health}),
ranging from very good to very bad, and activity limitation
(\texttt{hlthhmp}), which records whether respondents are hampered in daily
activities by longstanding illness, disability, infirmity, or a mental-health
problem. Both measures are harmonized to $[0,1]$, with higher values indicating
greater adversity. On the HFGI side, we use the online expression of bodily
pain (\texttt{pain}) and health-related limitation (\texttt{healthlim}). All
four indicators are standardized within the corresponding analytical sample.

The SEM likewise uses precomputed ESS country--year
aggregates. For Rounds 5--8, country--year means were calculated using
$\texttt{pspwght}\times\texttt{pweight}$, where \texttt{pspwght} is the
post-stratification weight incorporating the sampling-design weight and
\texttt{pweight} is the population-size weight. For Rounds 9--11, we used the
supplied ESS analysis weight, \texttt{anweight}. These weights were applied
when constructing the country--year aggregates but were not applied again
during SEM estimation.

The HFGI and ESS indicators are compared at the level of their
source-specific physical-adversity constructs rather than as one-to-one item
matches. The ESS core contains no direct measure of bodily pain, and
\texttt{hlthhmp} may include limitations caused by mental-health conditions.
Table~\ref{tab:hfgi-survey-crosswalk} summarizes the indicator blocks used in
the three domain-specific SEMs and clarifies the nature of their cross-source
correspondence.

\subsection{Aligning HFGI with WVS and ESS Data}

{For both survey sources, observations are matched to HFGI by
country and exact calendar year. A survey observation therefore enters an
analysis only when the corresponding annual HFGI indicators are available for
the same country and year.}

WVS observations were matched to HFGI observations by country and exact
calendar year, using the recorded year of each national WVS survey.
{The resulting HFGI--WVS dataset} contains 42 country--year
observations from 30 countries, distributed across ten survey years between
2012 and 2023. Because countries are not surveyed annually and different
respondents are sampled in each WVS wave, this dataset is an unbalanced panel
of repeated cross-sectional country-level aggregates rather than an
individual longitudinal panel.

Summing the total HFGI tweet count once for each matched country--year yields
an underlying analytical volume of 
approximately 277
million tweets.
Indicator-specific estimates are constructed from the subsets of tweets
classified as valid for the corresponding dimensions.

{The HFGI--ESS match follows the same country--year principle.
The resulting dataset contains 221 country--year observations for 32 countries
over 2011--2023 and covers ESS Rounds 5--11. Because ESS participation varies
across countries and rounds, these observations likewise form an unbalanced
panel of repeated cross-sectional country-level aggregates.}

For the ESS, calendar year is defined by the respondent's
interview-start year (\texttt{inwyys}), rather than by the nominal ESS round.
When a country's fieldwork spans two calendar years, respondents are assigned
to separate country--year aggregates according to their interview-start year.
If more than one ESS round contributes observations to the same country and
calendar year, the resulting estimates are pooled using their survey-weight
sums before being matched to the corresponding annual HFGI observation.

{For each survey source,} we use two representations of the
matched data. The country--year specification retains
{all available matched observations (42 for WVS and 221 for
ESS). The corresponding country-average specifications} first average the
available matched country--year observations within each country, yielding
{30 observations for WVS and 32 for ESS.} Each observed
indicator is standardized separately within the relevant analytical dataset
before SEM estimation. Accordingly, the variables entering the SEM have mean
zero and unit standard deviation rather than retaining the original HFGI
$[-1,1]$ or {survey} $[0,1]$ scales.

{Table~\ref{tab:hfgi-survey-crosswalk} presents the measurement
blocks used in both the HFGI--WVS and HFGI--ESS analyses.} Indicators appearing
in the same row jointly measure a source-specific latent construct; they should
not necessarily be interpreted as one-to-one equivalents. The final column
therefore distinguishes close conceptual correspondence from proxy and
cross-construct comparisons.

\subsection{Interpretation of the matched data}

{The HFGI and the WVS/ESS survey measures should therefore be
understood as distinct measurement systems. WVS and ESS indicators summarize}
survey-reported experiences and attitudes, whereas HFGI indicators capture
publicly expressed content identified in social-media text. Cross-national
differences in HFGI scores may reflect both differences in the underlying
constructs and differences in who uses Twitter, who enables geolocation, and
what users are willing to express publicly. Country--year matching improves
temporal comparability but does not make the populations represented by the
{different sources} equivalent. Accordingly, the SEM results in
the next section measure aggregate correspondence between survey-reported and
online-expressed constructs; they should not be interpreted as establishing
population representativeness, measurement equivalence, or causal
relationships.

\begin{table}[htbp]
\centering
\caption{{Crosswalk between HFGI and survey indicators used in
the structural equation models}}
\label{tab:hfgi-survey-crosswalk}
\footnotesize
\begin{tabularx}{\textwidth}{
  @{}
  >{\raggedright\arraybackslash}p{0.16\textwidth}
  >{\raggedright\arraybackslash}X
  >{\raggedright\arraybackslash}p{0.07\textwidth}
  >{\raggedright\arraybackslash}X
  >{\raggedright\arraybackslash}p{0.21\textwidth}
  @{}
}
\toprule
SEM construct &
HFGI indicators &
{Survey} &
{Survey indicators} &
Nature of the correspondence \\
\midrule

Mental and subjective well-being &
Happiness; life satisfaction; inner peace; depression;
anxiety &
{WVS} &
Happiness; life satisfaction; perceived choice and control &
Happiness and life satisfaction are direct conceptual
matches. Inner peace, depression, and anxiety broaden the HFGI factor beyond
the narrower WVS construct, while perceived choice and control captures agency
and has no direct HFGI counterpart in this model. \\

{Physical well-being} &
{Pain; health limitations} &
{ESS} &
{Self-rated health; activity limitation} &
{Domain-level correspondence. ESS provides a multi-indicator
physical-adversity construct, although there is no direct bodily-pain item and
activity limitation may also arise from mental-health problems.} \\

Spiritual and religious expression &
Belief in God; belief in an afterlife; religious comfort &
{WVS} &
Belief in an afterlife; importance of God; 
religious identity &
Close domain-level correspondence. Belief in an afterlife is
a direct match. HFGI belief in God and WVS importance of God are closely
related but not identical; the remaining indicators capture distinct aspects
of religious comfort and identity. \\

\bottomrule
\end{tabularx}

\vspace{0.5em}
\begin{minipage}{\textwidth}
\footnotesize
\textit{Note:} All indicators enter the SEM as standardized
country-level scores. HFGI depression, anxiety, pain, and health limitation,
together with ESS self-rated poor health and activity limitation, are oriented
so that higher values indicate greater adversity. The three domain models are
complementary analyses and do not assume measurement equivalence between HFGI
and the survey sources.
\end{minipage}
\end{table}

\subsection{SEM Analysis}

The structural equation modeling (SEM) analysis operationalizes the distinction between within-source
\emph{coherence} and between-source \emph{correspondence} introduced in Section~\ref{sec:framework}.
For each well-being domain, HFGI indicators define a latent factor representing
online expression, while the corresponding survey indicators define a separate
latent factor representing survey-reported experience or attitudes. The
measurement loadings describe the extent to which indicators within each
source cohere around their source-specific latent construct. The association
between the two latent factors, on the other hand, measures cross-source correspondence,
that is, the extent to which HFGI and survey data rank countries similarly
within the same domain.

All observed indicators are standardized within the analytical sample before
estimation. The two source-specific latent factors are estimated jointly using
robust maximum likelihood and are allowed to correlate.
For the country--year specifications, standard errors and
test statistics were made robust to repeated observations by clustering at the
country level. The country-average specifications contain one observation per
country and therefore do not require a cluster correction.
The reported 95\% confidence intervals for latent correlations
are Wald intervals based on robust standard errors; for the country--year
models, these standard errors incorporate the country-level cluster
correction.
Our focal parameter is
therefore the standardized latent correlation between the HFGI and survey
factors. A positive correlation indicates that countries scoring higher on the
latent construct in HFGI also tend to score higher on the corresponding
survey-based construct, whereas a weak correlation indicates that the two
measurement systems capture different cross-national variation.

We apply this framework to three domain-level models: \textit{mental and subject
well-being} and \textit{spiritual and religious expression} using the World Values Survey, and \textit{physical
well-being} using the European Social Survey. The indicators entering each
source-specific factor need not be item-by-item equivalents; the models assess
correspondence between conceptually related latent constructs rather than
measurement invariance or equivalence across data sources. The specific
measurement structure and any identifying constraints required by individual
models are described with the corresponding results.

Model adequacy is evaluated using parameter estimates, solution admissibility, and
standard global fit indices. We report the Comparative Fit Index (CFI), which
compares the fitted model with a null model; the Root Mean Square Error of
Approximation (RMSEA), which measures approximate model misfit per degree of
freedom; and the Standardized Root Mean Square Residual (SRMR), which summarizes
the average discrepancy between observed and model-implied standardized
covariances. As conventional guidelines, CFI values of approximately $.95$ or
higher, RMSEA values of approximately $.06$ or lower, and SRMR values of
approximately $.08$ or lower are generally indicative of good fit
\citep{hu1999cutoff}. These thresholds are treated as descriptive guidelines
rather than strict acceptance criteria. In particular, RMSEA can behave
poorly in models with few degrees of freedom or small samples. We therefore
interpret global fit jointly with convergence, the absence of inadmissible
estimates such as negative residual variances, the magnitude and stability of
factor loadings, and the substantive plausibility of the estimated latent
correlations. 

\section{Results} \label{sec:results}
The analyses examined three domains across two survey
sources. The HFGI--WVS analyses cover mental and subjective well-being and
spiritual and religious expression, using 30 countries and 42 matched
country--year observations. The HFGI--ESS analysis covers physical adversity,
using 32 countries and 221 matched country--year observations. For each
domain, separate HFGI and survey latent factors were estimated. Standardized
factor loadings describe coherence within each source, whereas the latent
correlation between the two source-specific factors describes cross-source
correspondence.

\subsection{Mental and subjective well-being} \label{sec:results-mental} 
\begin{figure}[H] \centering \includegraphics[ width=\textwidth ]{ 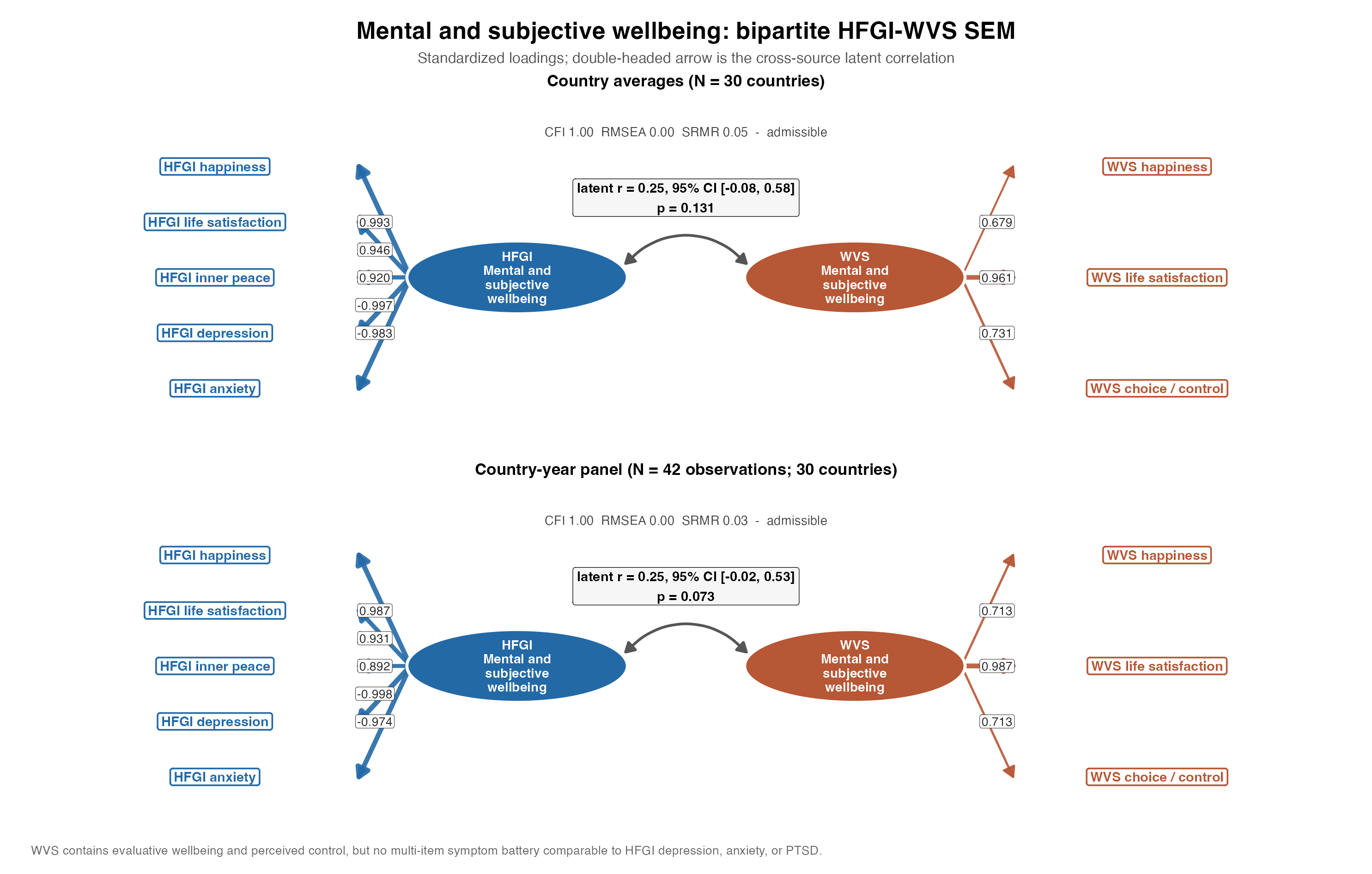 } \caption{ Bipartite HFGI-WVS measurement models for mental and subjective well-being. The upper panel presents the country-average model for 30 countries. The lower panel presents the country-year model based on 42 observations from 30 countries. Values adjacent to the measurement paths are standardized factor loadings. The double-headed arrow reports the standardized latent correlation between the HFGI and WVS factors. Depression and anxiety are inversely oriented relative to positive mental and subjective well-being. The WVS factor represents evaluative well-being and perceived choice and control; WVS does not contain a multi-item symptom battery directly comparable to the HFGI depression and anxiety indicators. } \label{fig:mental-sem} \end{figure}

The indicators of mental and subjective well-being loaded strongly on their respective source-specific factors (Figure~\ref{fig:mental-sem}). In the country-average model, the standardized HFGI loadings were .993 for happiness, .946 for life satisfaction, .920 for inner peace, $-.997$ for depression, and $-.983$ for anxiety. The negative loadings for depression and anxiety were consistent with the inverse orientation of these indicators relative to positive mental and subjective well-being.

The corresponding HFGI loadings in the country--year model were similar: .987 for happiness, .931 for life satisfaction, .892 for inner peace, $-.998$ for depression, and $-.974$ for anxiety. All corresponding residual variances remained positive. The smallest standardized residual variances were .005 for depression in the country-average model and .003 in the country--year model, indicating near-boundary, but not improper, solutions. The similarity of the loading patterns across the country-average and country--year specifications indicates that the five HFGI indicators formed a highly coherent mental and subjective well-being factor.

The WVS indicators also formed a coherent source-specific factor, although the loadings were more heterogeneous than those for HFGI. In the country-average model, the standardized loadings were .68 for happiness, .96 for life satisfaction, and .73 for choice and control. In the country-year model, the corresponding loadings were .71, .99, and .71. Life satisfaction was therefore the strongest WVS indicator in both specifications, while happiness and choice and control made substantial but comparatively smaller contributions. 

The reported global fit indices were favorable for both mental and subjective well-being models. The country-average model had CFI $=1.00$, RMSEA $=.00$, and SRMR $=.05$. The country-year model had CFI $=1.00$, RMSEA $=.00$, and SRMR $=.03$. Both solutions were classified as admissible. Given the small number of countries and the structured nature of the measurement models, however, these fit indices should be interpreted together with the parameter estimates and should not be treated as evidence that the HFGI and WVS measurement systems are equivalent. 

Despite strong coherence within each source, correspondence between the HFGI and WVS latent factors was modest. In the country-average model, the cross-source latent correlation was positive but statistically uncertain, $r=.25$, $p=.131$, 95\% CI $[-.08,.58]$. The country-year model produced the same estimated correlation, $r=.25$, with a smaller $p$-value, although the association did not reach conventional statistical significance: \(p=.073\), 95\% CI $[-.02,.53]$. 

The nearly identical latent correlations indicate that retaining the available country-year observations rather than averaging observations within countries did not materially alter the estimated magnitude of cross-source correspondence. The results therefore distinguish strong within-source measurement coherence from comparatively weak between-source correspondence. HFGI and WVS each produced internally coherent representations of mental and subjective well-being, but countries' relative positions on the two latent factors were only modestly aligned. 

\subsection{Spiritual and religious expression} \label{sec:results-spiritual} 
\begin{figure}[H] \centering \includegraphics[ width=\textwidth ]{ 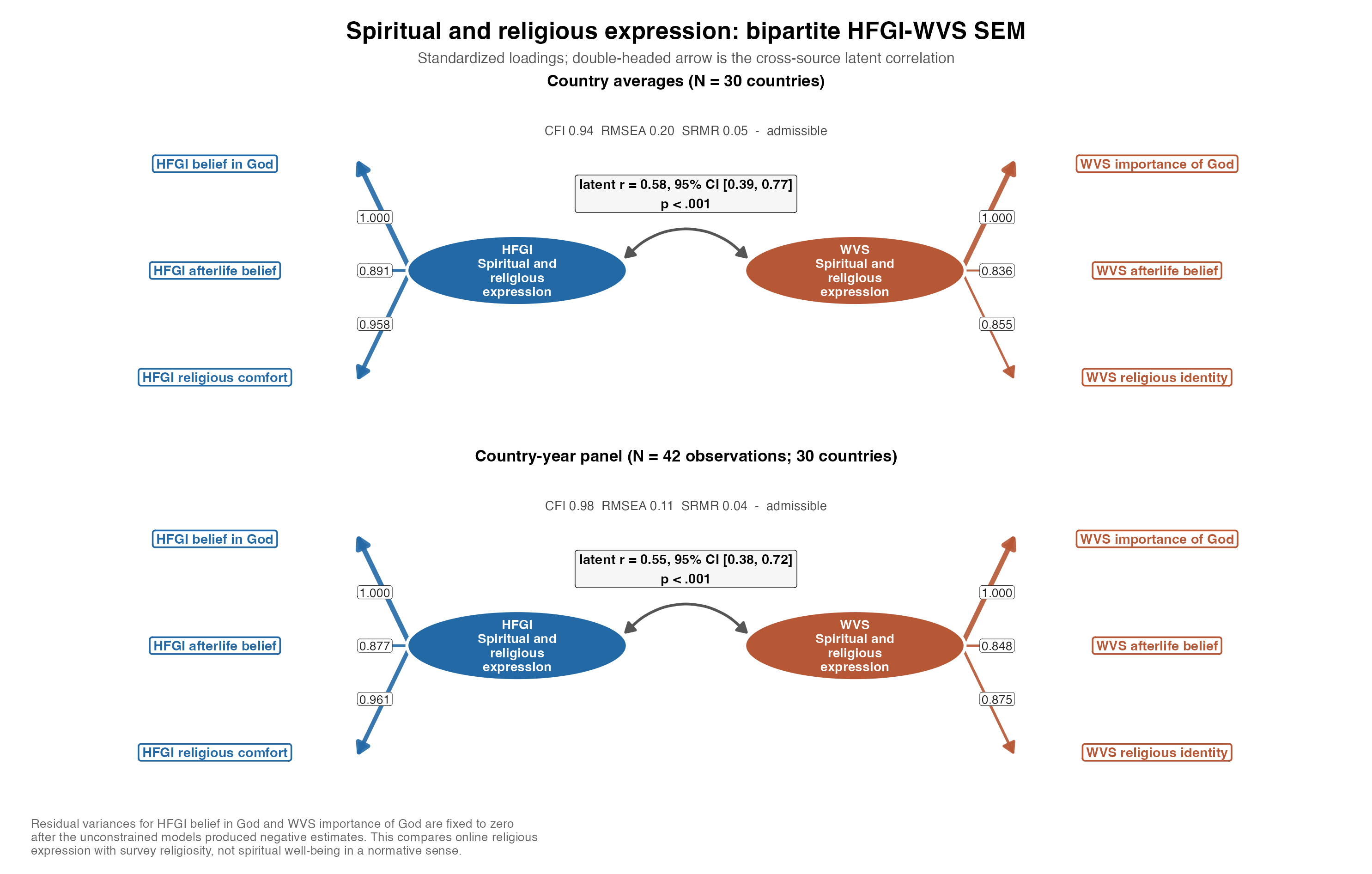 } \caption{ Bipartite HFGI-WVS measurement models for spiritual and religious expression. The upper panel presents the country-average model for 30 countries. The lower panel presents the country-year model based on 42 observations from 30 countries. Values adjacent to the measurement paths are standardized factor loadings. The double-headed arrow reports the standardized latent correlation between the HFGI and WVS factors. HFGI represents online religious expression through belief in God, afterlife belief, and religious comfort. WVS represents survey religiosity through the importance of God, afterlife belief, and religious identity. The residual variances of HFGI belief in God and WVS importance of God are fixed to zero after the unconstrained models produced negative estimates; their standardized loadings of 1.00 therefore reflect boundary constraints. The models assess cross-source correspondence in religious expression and should not be interpreted as measures of spiritual well-being in a normative sense.}  \label{fig:spiritual-sem} \end{figure}

The spiritual and religious expression indicators loaded strongly on their respective source-specific factors (Figure~\ref{fig:spiritual-sem}). The model included three HFGI indicators and three WVS indicators. The HFGI factor was measured by belief in God, afterlife belief, and religious comfort. The WVS factor was measured by the importance of God, afterlife belief, and religious identity. 

In the unconstrained models, the residual variances of HFGI
belief in God and WVS importance of God were negative in both analytical
samples. In the reported boundary-constrained specification, these two
residual variances were therefore fixed to zero. Their standardized loadings
are consequently equal to 1.00 by construction and should not be interpreted
as freely estimated evidence of perfect measurement.

In the country-average model, the standardized HFGI loadings were 1.00 for belief in God, .89 for afterlife belief, and .96 for religious comfort. The corresponding WVS loadings were 1.00 for the importance of God, .84 for afterlife belief, and .86 for religious identity. The four freely estimated loadings were therefore substantial, while the two loadings of 1.00 reflect the boundary constraints described above. Within HFGI, afterlife belief loaded less strongly than religious comfort. Within WVS, afterlife belief and religious identity both loaded substantially on survey religiosity.

The measurement pattern was highly similar in the country-year model. The standardized HFGI loadings were 1.00 for belief in God, .88 for afterlife belief, and .96 for religious comfort. The WVS loadings were 1.00 for the importance of God, .85 for afterlife belief, and .88 for religious identity. The close correspondence between the loading patterns in the country-average and country-year models indicates that the source-specific measurement structure was descriptively stable across the two analytical specifications. 

Cross-source correspondence was positive, moderate, and statistically significant in both models. In the country-average specification, the standardized latent correlation between HFGI spiritual and religious expression and WVS survey religiosity was $r=.58$, $p<.001$, 95\% CI $[.39,.77]$. In the country-year specification, the corresponding association was slightly smaller but remained similar in magnitude, $r=.55$, $p<.001$, 95\% CI $[.38,.72]$. Countries with higher levels of online religious expression therefore tended also to report greater religious importance, afterlife belief, and religious identity in the WVS. 

The similarity of the two estimates suggests that the principal cross-source association was not materially altered when the available country-year observations were retained rather than aggregated to one record per country. The estimates indicate a shared country-level religiosity gradient across online expression and survey reporting. Nevertheless, correlations of .58 and .55 represent moderate correspondence rather than equivalence. The two sources capture related but noninterchangeable representations of religious expression and religiosity. 

The global fit indices are mixed. For the country-average model, robust CFI was .94, robust RMSEA was .20, and SRMR was .05. The low SRMR indicates relatively small average standardized residual discrepancies, and the CFI approaches a conventionally favorable range. However, the value of RMSEA indicates that the model did not reproduce the observed covariance structure uniformly well. 

The country-year model showed comparatively stronger fit, with robust CFI $=.98$, robust RMSEA $=.11$, and SRMR $=.04$. The CFI and SRMR were favorable, while RMSEA remained somewhat elevated. Thus, the country-year specification provided a better approximation to the observed association structure than the country-average model, but support for overall model fit was not uniform across indices. Both models were numerically admissible. 

Overall, the positive latent correlations indicate that the two measurement systems capture a shared country-level pattern, while the remaining global misfit and differences in indicator content caution against treating the two factors as measurements of an identical construct. 

\subsection{Physical well-being}
\label{sec:results-physical}

\begin{figure}[H]
\centering
\includegraphics[width=\textwidth]{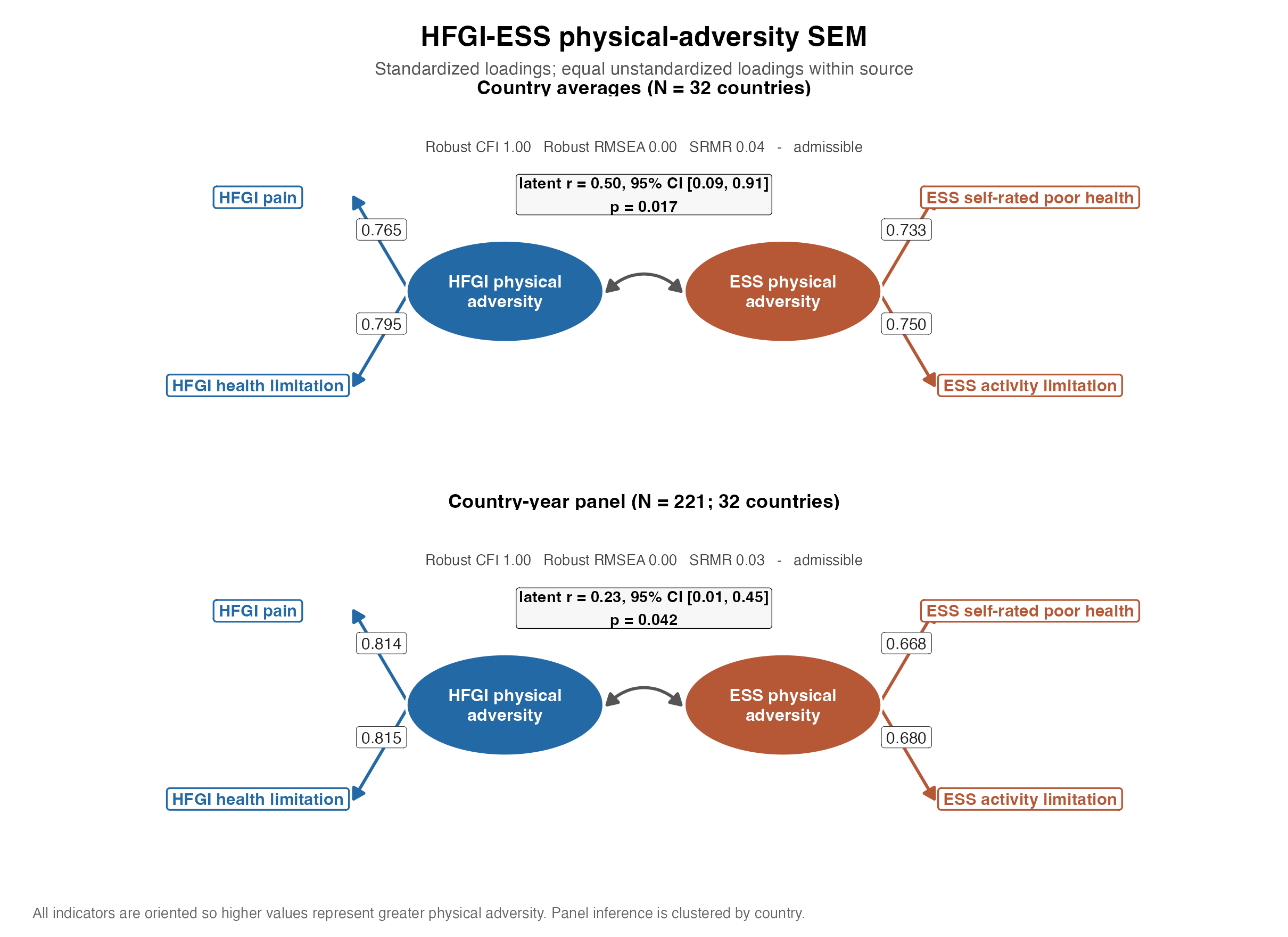}
\caption{
Bipartite HFGI--ESS measurement models for physical adversity. The upper panel
presents the country-average model for 32 countries. The lower panel presents
the country-year model based on 221 observations from 32 countries. Values
adjacent to the measurement paths are standardized factor loadings; the
equal-loading constraint applies to the corresponding unstandardized loadings
within each source. The double-headed arrow reports the standardized latent
correlation between the HFGI and ESS factors. All indicators are oriented so
that higher values denote greater physical adversity. Inference for the
country-year model is clustered by country.
}
\label{fig:physical-sem}
\end{figure}

The physical well-being analysis uses the ESS to provide a multi-indicator
survey comparison for this domain (Figure~\ref{fig:physical-sem}). Because all
four indicators are oriented toward adversity, the latent factors are
interpreted as physical-adversity factors: higher values correspond to greater
online expression of pain and health limitation in HFGI and poorer health and
greater activity limitation in the ESS.

Because each physical-adversity factor is based on only two indicators, we set the two unstandardized loadings within each source equal to obtain a stable model. This constraint is a modeling assumption rather than an empirical finding. The corresponding standardized loadings may still differ. We also estimated a model without the equality constraint as a sensitivity check.
 For the country-average data, the unconstrained model
converged, was admissible, and produced a similar latent correlation of
approximately $.47$. In the country-year panel, however, the unconstrained
model produced one negative residual variance and failed the post-estimation
admissibility check. Its coefficients are therefore not interpreted.

The indicators loaded substantially on their respective source-specific
factors. In the country-average model, the standardized HFGI loadings were
$.77$ for pain and $.80$ for health limitation. The corresponding ESS loadings
were $.73$ for poorer self-rated health and $.75$ for activity limitation. In
the country-year model, the HFGI loadings were $.81$ for both pain and health
limitation, while the ESS loadings were $.67$ for poorer self-rated health and
$.68$ for activity limitation. The similar loading patterns across the two
specifications indicate coherent source-specific representations of physical
adversity.

The country-average model produced a moderate positive point
estimate for the standardized latent correlation between the HFGI and ESS
physical-adversity factors, $r=.50$, $p=.017$ (95\% CI: $.09$--$.91$), but the
wide confidence interval indicates substantial uncertainty. The country-year model produced a weaker but still positive
association, $r=.23$, $p=.042$ (95\% CI: $.01$--$.45$). Countries with greater
online expression of pain and health limitation therefore also tended to
report, to some extent, poorer general health and greater activity limitation in the ESS.

The reported global fit indices were favorable for both physical-adversity
models. The country-average model had robust CFI $=1.00$, robust RMSEA $=.00$,
and SRMR $=.037$. The country-year model had robust CFI $=1.00$, robust
RMSEA $=.00$, and SRMR $=.032$. Both preferred solutions converged, passed the
post-estimation admissibility check, and contained no negative residual
variances. Because the equal-loading model has only three degrees of freedom,
however, these apparently excellent global fit statistics should be interpreted
as diagnostics rather than as strong evidence of exact fit or measurement
equivalence.

The lower correlation in the country--year specification
indicates that the estimated magnitude is sensitive to whether repeated annual
observations are retained or averaged within countries. Because the
country--year model combines within-country and between-country variation, this
contrast cannot identify whether correspondence is stronger for persistent
cross-national differences than for annual change.

As in the other domains, correspondence does not imply measurement
equivalence. The ESS contains no direct bodily-pain item in its cross-round
core, and the activity-limitation item can include limitations caused by
mental-health problems. The two source-specific factors therefore capture
related manifestations of physical adversity rather than identical
measurements.

Table~\ref{tab:sem-summary} summarizes the principal SEM results across the
three domains and the two analytical specifications. The country-average
models contain one observation per country, whereas the country--year models
retain all matched annual observations.

\begin{table}[H]
\centering
\caption{Summary of cross-source latent correspondence and model fit}
\label{tab:sem-summary}
\footnotesize
\setlength{\tabcolsep}{3.5pt}

\begin{tabularx}{\textwidth}{
@{}
l
l
l
r
r
r
c
r
r
r
>{\raggedright\arraybackslash}X
@{}
}
\toprule
Domain &
Source &
Analysis &
Countries &
\shortstack{Country-\\years} &
$r$ &
95\% CI &
CFI &
RMSEA &
SRMR &
Constraints \\
\midrule

Mental and \\ subj. well-being &
WVS &
Average &
30 &
-- &
.255 &
$[-.076,.585]$ &
1.000 &
.000 &
.053 &
None \\

 &
WVS &
Panel &
30 &
42 &
.254 &
$[-.024,.532]$ &
1.000 &
.000 &
.031 &
None \\

\addlinespace

Spiritual and \\ religious expression &
WVS &
Average &
30 &
-- &
.582 &
$[.393,.772]$ &
.945 &
.199 &
.047 &
Two residual variances fixed to zero \\

 &
WVS &
Panel &
30 &
42 &
.554 &
$[.384,.724]$ &
.983 &
.110 &
.039 &
Two residual variances fixed to zero \\

\addlinespace

Physical adversity &
ESS &
Average &
32 &
-- &
.499 &
$[.090,.909]$ &
1.000 &
.000 &
.037 &
Equal loadings within source \\

 &
ESS &
Panel &
32 &
221 &
.229 &
$[.009,.450]$ &
1.000 &
.000 &
.032 &
Equal loadings within source \\

\bottomrule
\end{tabularx}

\begin{minipage}{0.98\textwidth}
\footnotesize
\textit{Note.}
$r$ is the standardized correlation between the HFGI and survey latent
factors. Average models contain one observation per country; consequently,
country--years are reported only for the panel specifications. CFI and RMSEA
are robust fit indices. In the spiritual models, the residual variances of
HFGI belief in God and WVS importance of God were fixed to zero after the
unconstrained models produced negative estimates. In the physical-adversity
models, the two unstandardized loadings within each source were constrained
to equality. Inference for the country--year models accounts for clustering
by country.
\end{minipage}
\end{table}

The summary reveals distinct patterns across domains. Spiritual and religious
expression shows the strongest and most stable cross-source correspondence.
Mental and subjective well-being shows a smaller positive association, with
confidence intervals that include zero in both specifications. Physical
adversity produces a moderate country-average point estimate, although with
substantial uncertainty, and a smaller positive association in the
country--year model. These comparisons are descriptive and should not be
interpreted as a formal test of whether conceptual alignment determines the
strength of cross-source correspondence.

\paragraph{Supplementary country profiles.}
Figures~\ref{fig:supp-mental-heatmap}--\ref{fig:supp-physical-heatmap}
provide descriptive country-profile heatmaps for the indicators entering each
domain model. The hierarchically ordered profiles illustrate the distinction
between within-source coherence and cross-source correspondence at the country
level. They show comparatively broad cross-source alignment for spiritual and
religious expression, more frequent source-specific differences for mental and
subjective well-being, and partial alignment for physical adversity. Because
the ordering is derived from the combined HFGI--survey indicator space, the
dendrograms are exploratory and should not be interpreted as identifying
discrete country clusters or as an independent validation of either source.

\subsection{Country-influence sensitivity analyses}
Leave-one-country-out analyses assessed whether the estimated cross-source latent correlations were driven by individual countries (Table~\ref{tab:loo-combined}). The country-average and country--year models were re-estimated after sequentially omitting each country; in the country--year analyses, all observations belonging to the omitted country were removed together. The measurement specification was held fixed across refits. Influence was summarized by the range of admissible omission estimates and by $\Delta r=r_{(-c)}-r_{\mathrm{full}}$. {A refitted model was classified as admissible when it converged, passed the post-estimation checks, and produced no negative residual variances. Some omission models became inadmissible because removing a country reduced the already limited information available to estimate the latent-variable model, leading to an improper or numerically unstable solution.}

For mental and subjective well-being, the full-sample latent correlation was $r=.25$ in both specifications. The admissible leave-one-country-out estimates ranged from .19 to .41 in the country-average model and from .16 to .36 in the country--year model. The omission of Greece produced the largest changes, $|\Delta r|=.16$ and .10, respectively. Two of the 30 country-average omissions and two of the 30 country--year omissions produced inadmissible solutions. No admissible omission reversed the direction of the association, although conventional $p<.05$ status changed once in the country-average analysis and twice in the country--year analysis. The positive but modest correspondence was therefore directionally stable, while its precise magnitude and statistical uncertainty were somewhat sensitive to individual countries.  

Spiritual and religious expression showed the strongest country-level robustness. All 30 omission models were admissible in both analytical specifications, and the estimated latent correlation remained positive and statistically significant throughout. The leave-one-country-out estimates ranged from .55 to .65 around the country-average estimate of $r=.58$, and from .52 to .60 around the country--year estimate of $r=.55$. Omitting Iran produced the largest changes, but these were small, $|\Delta r|=.06$ and .05. There were no sign reversals or changes in conventional significance status. The moderate positive correspondence between online religious expression and WVS religiosity was therefore not attributable to any single country.  

For physical adversity, the leave-one-country-out estimates ranged from .40 to .59 around the country-average estimate of $r=.50$, and from .16 to .30 around the country--year estimate of $r=.23$. Omitting Ukraine produced the largest country-average change, $|\Delta r|=.10$, whereas omitting Germany produced the largest country--year change, $|\Delta r|=.07$. No admissible omission reversed the direction of the association. However, one country-average refit and five country--year refits were inadmissible, and conventional significance status changed in one and ten omissions, respectively. Thus, the positive direction of physical-adversity correspondence was stable among admissible refits, but the country--year inference was comparatively fragile and should be interpreted with particular caution.  

Across domains, the influence analyses reinforce the distinction between correspondence direction, magnitude, and statistical significance. Spiritual and religious expression displayed the most stable cross-source correspondence. Mental and subjective well-being remained positive but modest, with some sensitivity in magnitude. Physical adversity retained a positive direction, but the panel result showed the greatest sensitivity in admissibility and significance status. These diagnostics do not justify excluding influential countries; instead, they quantify how strongly the reported cross-source correspondence depends on particular national observations.

\begin{table}[H] \centering \caption{ Leave-one-country-out sensitivity of cross-source latent correlations across domains. } \label{tab:loo-combined}  \resizebox{\textwidth}{!}{%
\begin{tabular}{llcccccccc} \toprule Domain & Analysis & Full-sample $r$ & 95\% CI & $p$ & LOO range & Admissible/total & Sign reversals & $p$-status changes & Largest $|\Delta r|$ \\ \midrule Mental and \\ subj. well-being & Country average & .25 & $[-.08,.58]$ & .131 & .19--.41 & 28/30 & 0 & 1 & .16 (Greece) \\ & Country--year panel & .25 & $[-.02,.53]$ & .073 & .16--.36 & 28/30 & 0 & 2 & .10 (Greece) \\ 
\addlinespace 
Spiritual and \\ religious expression & Country average & .58 & $[.39,.77]$ & $<.001$ & .55--.65 & 30/30 & 0 & 0 & .06 (Iran) \\ & Country--year panel & .55 & $[.38,.72]$ & $<.001$ & .52--.60 & 30/30 & 0 & 0 & .05 (Iran) \\ 
\addlinespace 
Physical adversity \\ & Country average & .50 & $[.09,.91]$ & .017 & .40--.59 & 31/32 & 0 & 1 & .10 (Ukraine) \\ & Country--year panel & .23 & $[.01,.45]$ & .042 & .16--.30 & 27/32 & 0 & 10 & .07 (Germany) \\ \bottomrule \end{tabular}%
} 
\begin{minipage}{0.98\textwidth} \footnotesize \textit{Note.} LOO = leave-one-country-out. The LOO range is calculated from numerically admissible refits only. $\Delta r=r_{(-c)}-r_{\mathrm{full}}$. The country in parentheses produced the largest absolute change. Changes in whether $p<.05$ are reported separately from sign reversals. \end{minipage} \end{table}

\section{Summary and Discussion}

Across two independent survey sources and three substantive
domains, the HFGI shows meaningful but construct-specific correspondence with
survey measures. The observed ordering of associations is consistent with the
interpretation that closer conceptual alignment may accompany stronger
cross-source correspondence, while weaker correspondence appears where the
indicator sets capture different aspects of a broader domain. Because
conceptual alignment was not independently measured and only three domains were
compared, this pattern is descriptive and should not be interpreted as evidence
of a general relationship. It nevertheless provides a more informative basis
for interpreting the results than a simple distinction between ``agreement''
and ``disagreement.''

The contrast is clearest across domains. Spiritual and religious expression
shows the strongest correspondence ($r=.58$) {in the country-average model}, consistent with the relatively
close overlap between the HFGI and WVS indicators. Physical
well-being yielded a moderate country-average point estimate ($r=.50$), but the wide 95\% confidence interval ($.09$--$.91$) indicates substantial
uncertainty. The physical model was assessed with a different survey source and
with indicators that were related rather than identical. Mental and subjective well-being shows a
weaker association ($r=.25$), but here the two latent constructs differ more
substantially in content: HFGI includes inner peace, depression, and anxiety,
whereas the WVS factor is centered on happiness, life satisfaction, and
perceived choice and control. The weaker association therefore need not be
interpreted as a failure of {the HFGI measurement approach}, but as evidence that
social media indicators and surveys capture partly different dimensions of the broader well-being domain.

Indeed, near-perfect correspondence across all domains would have been
surprising. HFGI and survey measures are generated through fundamentally
different measurement processes. Surveys elicit standardized responses from
sampled respondents, whereas HFGI summarizes voluntarily expressed language
from geolocated social-media users. If the two systems produced virtually
identical country rankings regardless of domain or indicator content, this
would imply an unexpectedly strong equivalence between elicited self-report and
public expression. The fact that correspondence is positive but incomplete,
and varies in substantively interpretable ways across domains, is therefore
consistent with the idea that the two sources contain both shared signal and
source-specific information.

This interpretation also helps clarify what external validation should mean for
digital-trace measures. The objective need not be to reproduce survey rankings
perfectly. Rather, validation concerns whether independently generated
measurement systems recover related cross-national structure when they are
intended to capture related constructs, and whether departures from
correspondence can be understood from differences in construct definition and
measurement process. {In the present study, the observed
ordering of associations is consistent with, but does not establish, the
possibility that closer conceptual overlap is associated with stronger
cross-source convergence.}

HFGI should therefore be viewed as complementary to, rather than
interchangeable with, population surveys. Surveys provide standardized
population-based measures, while HFGI captures a denser record of public
expression across space and time. Their value in combination lies precisely in
the fact that they are not identical measurement systems: convergence provides
evidence of shared underlying structure, whereas divergence may reflect
dimensions of social experience that are expressed differently online and in
survey responses. The results of these three domains therefore support a view
of HFGI as a distinct but empirically connected measurement system whose degree
of convergence with surveys is construct and context dependent.

\section{Limitations}
\label{sec:limitations}

Several limitations qualify the interpretation of the results. The analyses are
based on a relatively small number of countries, particularly for the WVS
models, and the matched country-year data are unbalanced repeated
cross-sections rather than individual longitudinal panels. HFGI and survey
measures also represent different effective populations and different
measurement processes, and the indicators defining corresponding latent
constructs are conceptually related rather than identical. The results should
therefore be interpreted as evidence of cross-source correspondence at the
aggregate level, not as measurement equivalence or as estimates of individual
well-being.

\subsection{Potential Biases and Limitations of Social-Media Indicators}
\label{sec:issues}

The HFGI indicators describe expressions among geolocated Twitter users rather
than representative attitudes or experiences in the national population.
Participation on Twitter, the use of geolocation, and the decision to post
publicly are all selective processes. Twitter users differ systematically from
the wider population in age, education, income, and place of residence
\citep{wojcikhughes2019,mislove2011,hargittai2020potential}, while geolocated content is
particularly concentrated in urban areas \citep{malik2015}. Country-level
aggregation may reduce random variability, but it does not correct systematic
differences in platform use, geolocation, or demographic composition across
countries and years \citep{olteanu2019social}.

\subsection{Large Language Model Classification Error}

The HFGI indicators depend on the automated interpretation of short and often
informal social-media texts. Ambiguity, sarcasm, idiomatic language, and
culture-specific forms of expression may therefore introduce classification
error. Via fine-tuning and validation against manually annotated examples, we
improved alignment with the intended constructs but cannot fully eliminate
these sources of uncertainty \citep{carammia2026rethinking}.

\subsection{SEM Design and Statistical Inference}

The SEM results are descriptive and associational: the latent correlations
measure aggregate correspondence between online-expression and survey-report
factors and should not be interpreted causally. The relatively small number of
countries, especially in the WVS analyses, limits statistical precision and
requires caution in interpreting global fit indices. Country-year models
combine between-country and within-country variation and therefore do not
constitute pure longitudinal within-country analyses. For physical well-being,
the two-indicator factors also require equal-loading constraints in the
preferred specification; these constraints stabilize estimation but should be
viewed as substantive modeling assumptions rather than evidence of measurement
equivalence.
The near-unit HFGI mental-health loadings were accompanied by
small but positive residual variances, particularly for depression, and should
be interpreted as near-boundary estimates rather than as evidence of perfect
measurement.

\subsection{Cross-National and Multilingual Comparability}

Related validation using human-annotated posts from the GeoTweet Archive
provides encouraging evidence that fine-tuned language models can transfer
topic recognition across languages and that multilingual fine-tuning can reduce
language-related performance differences. See, e.g., \cite{nasuto2026language}:
although that study concerns immigration discourse, they empirically observed good performance on the HFGI indicators. 
Moreover, because our analysis uses
country-level aggregates, it cannot test measurement invariance within either
HFGI or the WVS and ESS measures. Country rankings may therefore combine
substantive differences with language-specific classifier performance, cultural
norms of public expression, survey translation, differences in question
interpretation, and cross-national response styles.


\section*{Data and Code Availability}
The analysis code and derived country--year datasets will be made available under the github repository
\url{https://github.com/siacus/} when in final stage. Only yearly country-level aggregates
are distributed. The analytical files do not contain WVS or ESS microdata,
tweet-level records, or user identifiers. 

\section*{Ethics Statement}

This study did not involve human participants or personal data. All analyses
were conducted using precomputed country--year indicators. The analytical
datasets contain no respondent-level records, social-media posts, user
identifiers, or other information that can be linked to individuals.

\section*{Statements and Declarations}
\paragraph{Competing Interest:} The authors declare no competing interest.

\bibliographystyle{apalike}
\bibliography{references}

\clearpage 

\appendix 
\section{Supplementary materials} \label{sec:supplementary-materials} 

\setcounter{figure}{0} \renewcommand{\thefigure}{S\arabic{figure}} \setcounter{table}{0} \renewcommand{\thetable}{S\arabic{table}}

\subsection{Supplementary country-profile analyses} \label{sec:supplementary-country-profiles} 
The supplementary heatmaps provide indicator-level descriptions of the country patterns underlying the domain-specific latent-variable models. For each domain, country-average indicators were standardized across the countries included in the relevant analysis. Countries were then ordered using Ward's minimum-variance hierarchical clustering based on Euclidean distances in the combined HFGI-survey indicator space. The heatmaps are descriptive rather than confirmatory. The dendrograms summarize relative similarities among the observed country profiles but do not establish discrete or stable country types. The resulting order can depend on the indicators included, their orientation, the treatment of missing data, the distance metric, and the country sample. 

\subsubsection*{Mental and subjective well-being} \label{sec:supp-mental-profiles} 
\begin{figure}[H] \centering \includegraphics[ width=\textwidth ]{ 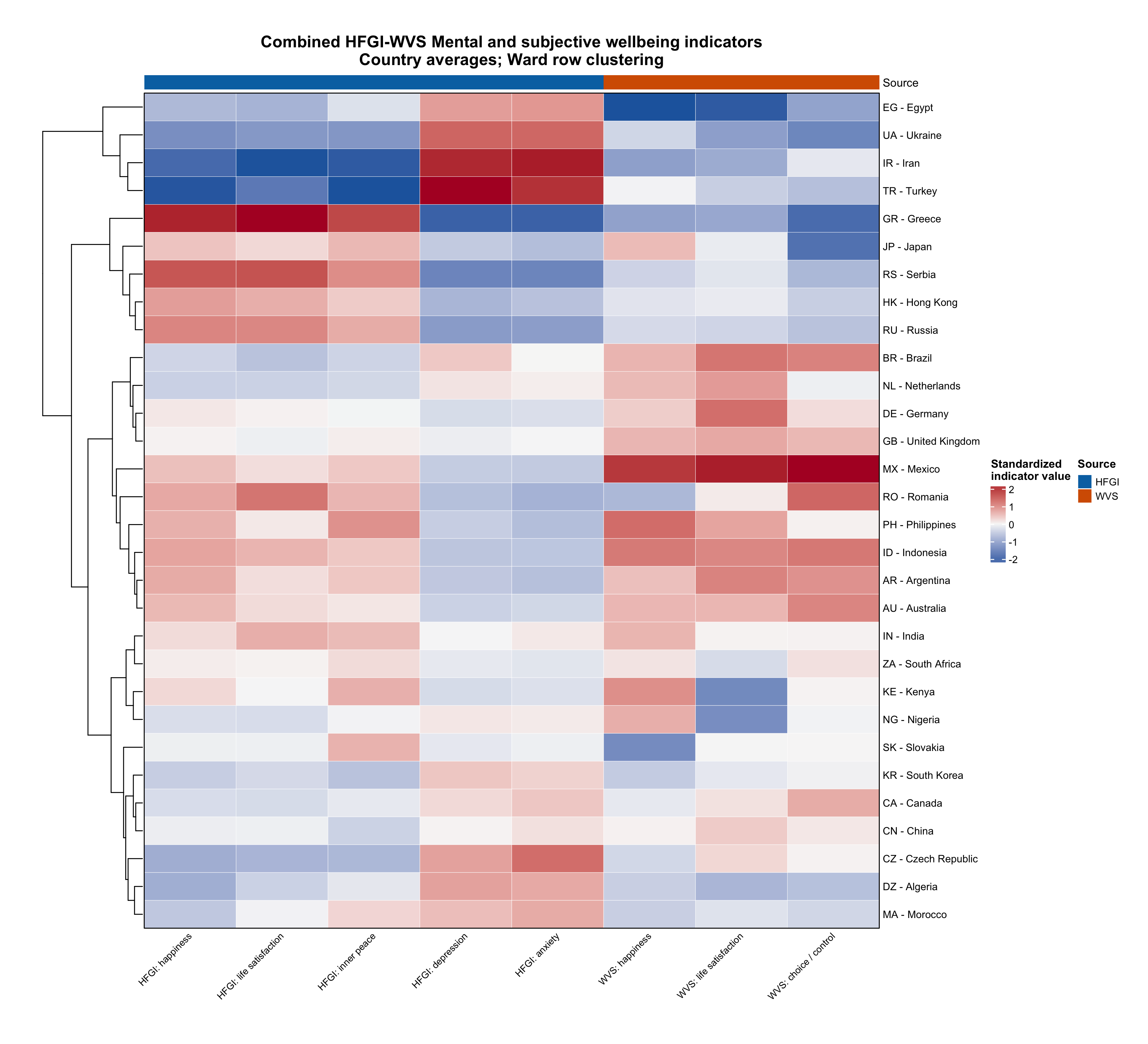 } \caption{ Combined HFGI and WVS country profiles for mental and subjective well-being. Cells contain indicator values standardized across countries. Red indicates a value above the indicator-specific cross-country mean, blue indicates a value below the mean, and white indicates a value near the mean. HFGI depression and anxiety retain their original symptom orientation, so higher values indicate greater depression or anxiety. Countries are ordered using Ward's minimum-variance hierarchical clustering based on Euclidean distances across the complete set of displayed indicators. The dendrogram is descriptive and does not impose discrete country clusters. } \label{fig:supp-mental-heatmap} \end{figure}

Figure~\ref{fig:supp-mental-heatmap} displays the combined HFGI and WVS indicators of mental and subjective well-being. The HFGI indicators comprise happiness, life satisfaction, inner peace, depression, and anxiety. The WVS indicators comprise happiness, life satisfaction, and perceived choice and control. 

The figure shows substantial coherence within each measurement source. HFGI happiness, life satisfaction, and inner peace generally move together, while depression and anxiety tend to vary in the opposite direction. The WVS indicators also show broadly similar country profiles across happiness, life satisfaction, and choice and control. These patterns are consistent with the strong source-specific factor loadings reported in the main analysis. 

Cross-source correspondence is less uniform. Iran and Turkey have low HFGI scores for happiness, life satisfaction, and inner peace and high HFGI depression and anxiety, while the corresponding WVS scores are closer to the cross-country average. Greece displays the reverse configuration, with particularly high HFGI positive well-being and low HFGI depression and anxiety but below-average WVS scores. Serbia, Hong Kong, and Russia also have positive HFGI profiles accompanied by comparatively lower WVS scores. Other countries show greater cross-source agreement. Brazil, Mexico, the Philippines, Indonesia, and Australia generally have positive or near-positive positions on both the HFGI positive well-being indicators and the WVS well-being indicators. Egypt and Ukraine tend to occupy the lower part of both source-specific distributions, although the magnitude differs across individual indicators.

The mixture of convergent and divergent country profiles is consistent with the modest latent correlation reported in the main analysis. HFGI and WVS each organize countries coherently within their own indicator sets, but the two sources do not produce an identical country ranking. Divergent profiles are visible across several countries and in multiple parts of the country distribution.

\subsubsection*{Spiritual and religious expression} \label{sec:supp-religiosity-profiles} 
\begin{figure}[H] \centering \includegraphics[ width=\textwidth ]{ 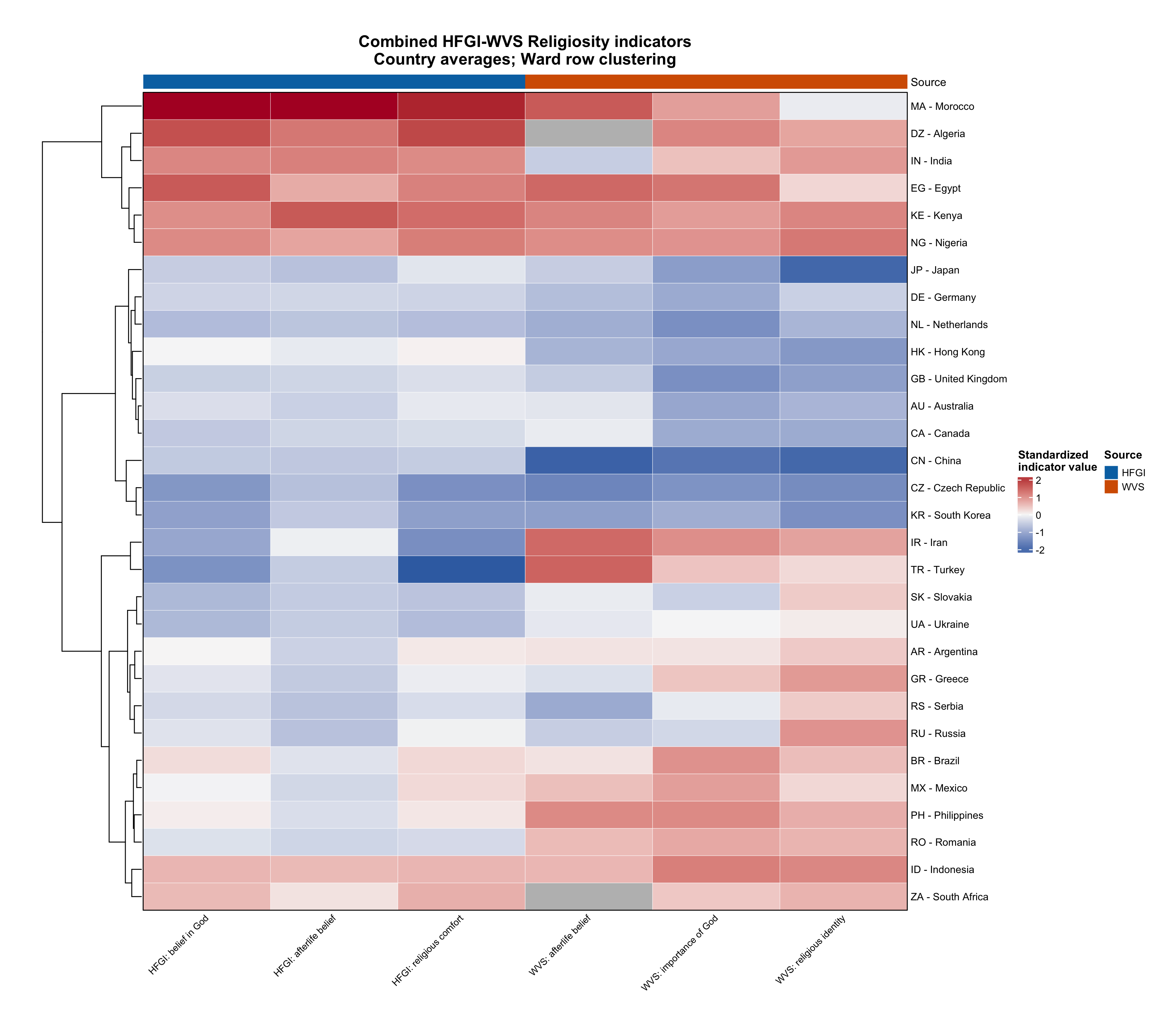 } \caption{ Combined HFGI and WVS country profiles for spiritual and religious expression. HFGI indicators comprise belief in God, afterlife belief, and religious comfort; WVS indicators comprise afterlife belief, the importance of God, and religious identity. Cells show indicator values standardized across countries. Red indicates values above the indicator-specific cross-country mean, blue indicates values below the mean, white indicates values near the mean, and gray indicates missing values. Countries are ordered using Ward's minimum-variance hierarchical clustering based on Euclidean distances across the six displayed indicators. Missing standardized values were replaced by zero only when calculating clustering distances. The dendrogram is descriptive and does not establish discrete country types. } \label{fig:supp-religiosity-heatmap} \end{figure}

Figure~\ref{fig:supp-religiosity-heatmap} presents the combined country profiles for the six indicators in the spiritual and religious expression model. The HFGI indicators are belief in God, afterlife belief, and religious comfort. The WVS indicators are afterlife belief, the importance of God, and religious identity. 

The heatmap shows a broad religiosity gradient across both sources. Morocco, Algeria, India, Egypt, Kenya, and Nigeria generally have above-average values across several HFGI and WVS indicators. Japan, Germany, the Netherlands, Hong Kong, the United Kingdom, Australia, Canada, China, the Czech Republic, and South Korea generally have below-average values across both sources. This shared pattern is consistent with the moderate positive latent correspondence reported in the main analysis. 

Several countries nevertheless display source-specific differences. Iran and Turkey have below-average HFGI religious-expression scores but above-average WVS scores for afterlife belief and the importance of God. Argentina, Greece, Russia, and Romania also tend to have more positive WVS than HFGI profiles. Indonesia shows comparatively positive values across both sources, while Brazil, Mexico, the Philippines, and South Africa exhibit more heterogeneous profiles. 

These patterns indicate that HFGI and WVS capture a common broad religiosity gradient but do not produce identical country rankings. The heatmap therefore complements the latent correlations, $r=.58$ in the country-average model and $r=.55$ in the country-year model, by showing where cross-source convergence and divergence occur at the indicator and country levels. The clustering is exploratory and should not be interpreted as a classification of discrete country types. 

\subsubsection*{Physical well-being} \label{sec:supp-physical-profiles} 
\begin{figure}[H] \centering \includegraphics[ width=0.88\textwidth ]{ 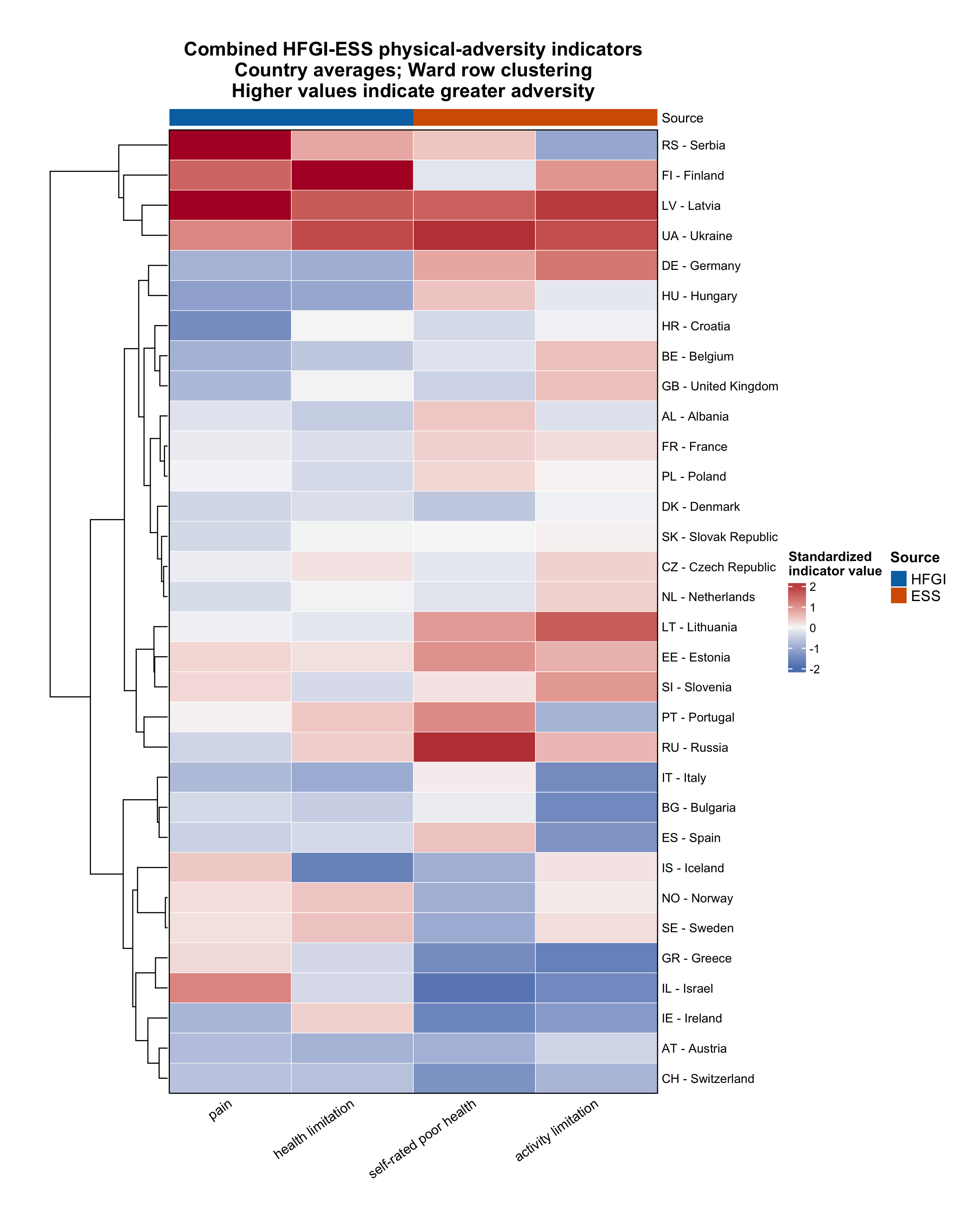 } \caption{ Combined HFGI and ESS country profiles for physical adversity. Cells contain indicator values standardized across countries. All indicators are oriented so that higher values represent greater physical adversity. Red therefore indicates greater pain, health limitation, poor self-rated health, or activity limitation relative to the cross-country mean; blue indicates lower physical adversity. Countries are ordered using Ward's minimum-variance hierarchical clustering based on Euclidean distances across the four displayed indicators. The dendrogram is descriptive and does not impose discrete country clusters. } \label{fig:supp-physical-heatmap} \end{figure}

Figure~\ref{fig:supp-physical-heatmap} presents the physical well-being analysis in adversity-oriented form. The HFGI indicators are pain and health limitation, while the ESS indicators are self-rated poor health and activity limitation. All four indicators are oriented so that higher values represent greater physical adversity. 

Several countries at the top of the heatmap show consistently high physical adversity. Latvia and Ukraine have above-average values on all four indicators. Finland also has high HFGI pain and health limitation, together with positive ESS activity limitation. Serbia has particularly high HFGI pain but a below-average ESS activity-limitation score, illustrating that the two sources do not always identify the same component of physical adversity. 

Several countries occupy an intermediate part of the distribution. Germany combines comparatively low HFGI pain and health limitation with above-average ESS poor health and activity limitation. Belgium and the United Kingdom display a similar, though less pronounced, source difference. Lithuania, Estonia, and Slovenia have relatively high ESS adversity despite HFGI values that are closer to the cross-country average. Italy, Bulgaria, Spain, Greece, Israel, Ireland, Austria, and Switzerland generally have below-average scores on the ESS indicators of poor health and activity limitation. HFGI profiles within this part of the heatmap are more heterogeneous. Israel, for example, has comparatively high HFGI pain but low ESS adversity, whereas Ireland has below-average HFGI pain and above-average HFGI health limitation while remaining low on both ESS indicators. 

The physical well-being heatmap therefore indicates partial rather than complete alignment between online and survey measures of physical adversity. Some countries have consistently high or low values across sources, whereas others differ according to whether physical well-being is represented through online references to pain and health limitation or through survey reports of poor health and activity restriction. 

\subsubsection*{Cross-domain interpretation} \label{sec:supp-cross-domain-profiles} 
Taken together, Figures~\ref{fig:supp-mental-heatmap}, \ref{fig:supp-religiosity-heatmap}, and \ref{fig:supp-physical-heatmap} reinforce the distinction between coherence within measurement systems and correspondence across measurement systems. 

Within each source, conceptually related indicators generally produce recognizable country profiles. However, the country rankings generated by online expression and survey reporting are not identical. The strongest visual cross-source alignment appears in the spiritual and religious expression domain, where both HFGI and WVS identify broad groups of countries with comparatively high or low religiosity. This pattern is consistent with the larger latent correlation reported for this domain. The mental and subjective well-being domain shows more frequent source-specific differences, consistent with its weaker latent correspondence. The physical-adversity analysis similarly shows both convergent country profiles and cases in which HFGI and ESS indicators point in different directions. 

The supplementary analyses do not establish why individual country profiles differ across sources. The patterns may reflect differences in indicator coverage, effective populations, temporal alignment, linguistic expression, public communication, survey elicitation, or the distinction between experiential, evaluative, behavioral, and symptom-based components of the domains. The heatmaps should therefore be interpreted as descriptive evidence about where the measurement systems converge and diverge, rather than as validation of one source against another.

\end{document}